\documentclass[aps,prl,twocolumn,showpacs,superscriptaddress,
amsmath,amssymb,floatfix]{revtex4}
\usepackage{graphicx}% Include figure files
\usepackage{dcolumn}% Align table columns on decimal point
\usepackage{bm}% bold math
\begin{document}
\title{Fluid adsorption near an apex: Covariance between complete
and critical wetting}
\author{A. O. Parry}
\affiliation{Department of Mathematics, Imperial College 180 Queen's Gate,
London SW7 2BZ, United Kingdom}
\author{M. J. Greenall}
\affiliation{Department of Mathematics, Imperial College 180 Queen's Gate,
London SW7 2BZ, United Kingdom}
\author{J.M. Romero-Enrique}
\altaffiliation
[On leave from ]{Departamento de F\'{\i}sica At\'omica, Molecular y 
Nuclear, Area de F\'{\i}sica Te\'orica, Universidad de Sevilla, 
Apartado de Correos 1065, 41080 Sevilla, Spain}
\affiliation{Department of Mathematics, Imperial College 180 Queen's Gate,
London SW7 2BZ, United Kingdom}
\begin{abstract}
Critical wetting is an elusive phenomenon for solid-fluid interfaces. Using
interfacial models we show that the diverging length scales, which characterize
complete wetting at an apex, precisely mimic critical wetting with the apex 
angle behaving as the contact angle. Transfer matrix, renormalization group 
(RG) and mean field analysis (MF) shows this covariance is obeyed in 2D, 
3D and for long and short ranged forces. This connection should be 
experimentally accesible and provides a means of checking theoretical 
predictions for critical wetting. 
\end{abstract}
\maketitle
Advances in the controlled fabrication of micropatterned 
substrates have stimulated the experimental and theoretical study of
fluid adsorption at tailored surfaces \cite{1,2,3,4,5}. For example, 
Mistura and co-workers \cite{5} have recently investigated complete
wetting of Ar on several parallel arrays of wedges \emph{and apexes}. 
They show convincingly that the adsorption within the (independent)
wedge regions is geometry dominated and distinct
from the planar complete wetting case. As well as having implications for 
microfluidics such studies have also revealed a number of unexpected 
results relating interfacial fluctuation effects and substrate geometry
which have wide application to other phase transitions (see later). 
Here we use effective Hamiltonian theory to show that \emph{complete} 
wetting on \emph{apex} shaped substrates reveals a hidden connection 
(covariance) with \emph{critical} (continuous) wetting transition occurring 
on planar surfaces \cite{6,7}. The covariance emerges when one considers how, 
at bulk coexistence, the mean height $l_A(\alpha)$ of the unbinding 
interface above the apex tip depends on the apex angle $\alpha$. For 
shallow apexes we show that $l_A(\alpha)$ is identical
to the mean interfacial thickness occurring at a particular class of
critical wetting transition with the apex angle playing the role of an
effective contact angle. The covariance is valid for 2D and 3D apexes, 
for arbitrary intermolecular forces and is, we believe, a general feature 
which should also be present in more microscopic models.

The central result of our article is the following covariance
relation for the interfacial height probability distribution function (PDF) 
which precisely quantifies the influence of the apex geometry on the complete
wetting film. We emphasize that the PDF contains a great deal of information
determining the (local) interfacial height, roughness and the
scaling properties of the density profile. Let $P_A^U(l,\alpha)$ denote
the PDF for the interface height above the apex where the superscript 
refers to the repulsive binding potential $U(l)\sim Bl^{-p}$ for a complete 
wetting film (see below). Let $P_{\pi}^V(l;\theta)$ denote the PDF for a 
planar critical wetting transition written in terms of the contact angle 
$\theta$. Here $V(l)$ is the binding potential for a critical wetting 
transition which, of necessity, contains attractive and repulsive terms. The 
covariance relation, valid for small $\alpha$ and at bulk coexistence, reads
\begin{equation}
P_A^U(l,\alpha)=P_{\pi}^V(l;\alpha)
\label{apexcov}
\end{equation}
where the covariant binding potential
\begin{equation}
V(l)=-Al^{-\chi/2}+U(l)
\label{vcov}
\end{equation}
with $\chi=\min (p,\tau)$ and $A$ determining the effective contact
angle (related to $\alpha$). Here $\tau=2(1-\zeta)/\zeta$ and 
$\zeta$ are the entropic repulsion and interfacial wandering 
exponents respectively \cite{8}. Thus the apex locally binds 
the complete wetting film and induces an effective attractive term in 
the binding potential, twice the range of the dominant intermolecular 
or entropic repulsive term. A similar rule applies in 3D for exponentially 
decaying potentials. Recall that in contrast to abundant experimental 
studies of complete wetting (including recent work on systems with
short ranged forces \cite{9}) critical wetting transitions are rather 
rare \cite{6,7} for which no examples are known currently for solid-fluid
interfaces. The covariance discussed here provides a means of effectively 
inducing critical wetting behaviour using complete wetting films. 

\begin{figure}
\includegraphics[width=8.6cm]{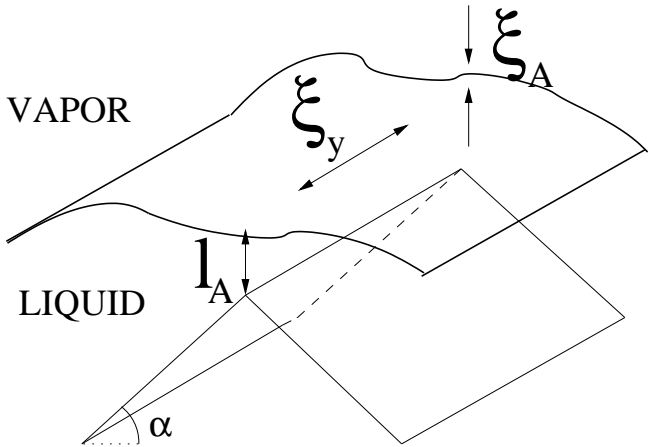}
\caption{Schematic illustration of 3D apex complete wetting, showing a section
of a typical interfacial configuration above the tip. Diverging length
scales are highlighted.\label{fig1}}
\end{figure}

Consider the interface between an infinite apex and a bulk vapor
 at temperature $T$ and chemical potential $\mu\le\mu_{sat}(T)$ (see Fig. 
\ref{fig1}). We suppose that the flat wall ($\alpha=0$) is completely 
wet by the liquid phase at coexistence $\delta \mu\equiv\mu_{sat}-\mu(T)=0^+$ 
corresponding to zero contact angle. The wall shape is described by a 
height function $z_A=-\tan\alpha\vert x \vert$ in the $(x,z)$ plane 
although we shall only be interested in the case of shallow apexes for 
which we may approximate $\tan\alpha\approx\alpha$. Macroscopically far 
from the apex tip the height of the interface above the wall is the same 
as that occurring for a flat wall. Since the liquid-vapor interface is
required to round the apex, surface tension restrictions imply
that the local height $l_A$ above the apex tip is smaller and remains finite
even in the limit of bulk coexistence. We wish to evaluate 
the mean interfacial height $l_A(\alpha)$ and interfacial roughness (r.m.s.
interfacial width) $\xi_A(\alpha)$ at bulk coexistence and the critical 
exponents
\begin{equation}
l_A(\alpha)\sim
\alpha^{-\beta_A},{\indent}\xi_A(\alpha)\sim\alpha^{-\nu_A}
\label{exponentS}
\end{equation}
For the 3D apex we also wish to determine the transverse
correlation length $\xi_y(\alpha)\sim\alpha^{-\nu_y}$ pertinent to
correlations along the apex. Correlations in the $x$ direction are not
described by a finite correlation length and fluctuations are not 
localised to a region near the apex.

We begin with the 2D apex. The starting point for our calculations is the 
interfacial Hamiltonian model 
\begin{equation}
\beta H_A[l] = \int dx \left\{ \frac{\Sigma}{2} \left(\frac {dl}{dx} \right)^2
+U(l+\alpha\vert x\vert) \right\}
\label{effapex}
\end{equation}
where $l(x)$ is the local height of the interface above the $z=0$
reference line, $\Sigma$ is the (reduced) stiffness coefficient (surface
tension) of the liquid-vapor interface and $U(l)$ denotes the binding 
potential modelling the complete wetting behavior pertinent to the planar 
system $\alpha=0$. For shallow apexes, it is permissable to assume 
the interface interaction with the wall 
occurs via the relative vertical height $\tilde l\equiv l+\alpha\vert x\vert$. 
The binding potential $U(l)$ has an infinite hard wall repulsion and decays as
\begin{equation}
U(l)= (\rho_l-\rho_g)\beta \delta \mu l+B l^{-p}
\label{binding}
\end{equation}
where, for the moment we have allowed for a finite bulk-order field
$\delta\mu>0$. Here $B$ is a positive Hamaker constant whilst
$p$ accounts for the range of the intermolecular forces. Exponentially
decaying binding potentials will also be considered in our discussion of 
the 3D apex. For dimensions in which the free liquid-vapor 
interface is rough, critical effects at planar complete 
wetting transitions fall into two classes \cite{8,10}:
a MF regime for $p<\tau$ and a fluctuation-dominated regime for
$p>\tau$. Heuristically this arises from the interplay between the direct
intemolecular repulsion $\sim l^{-p}$ and the effective entropic replusion
$U_{fl}(l)\sim l^{-\tau}$. In the present paper we restrict our attention
to pure systems for which the interface is rough for $d\le 3$ and
$\zeta=(3-d)/2$. For fixed $p$ the upper critical dimension for complete 
wetting is $d_{co}=3-4/(p+2)$ \cite{10}.

The model can be studied using
transfer matrix methods previously developed for the wedge geometry \cite{11}.
Care must be taken in defining an infinite apex geometry and it is
convenient to first consider a finite apex extending over the range 
$[-L/2,L/2]$ and impose periodic boundary conditions at the edges. After
taking the thermodynamic limit $L\to\infty$ at finite $\delta\mu >0$ it is
straightforward to derive an expression for the interfacial height 
probability distribution function at arbitrary position $x$ along the wall. 
At the apex mid-point symmetry considerations simplify this expression 
considerably 
\begin{equation}
P_A^U(l;\alpha)\propto \vert\psi_0(l)\vert ^2 e^{-2\Sigma \alpha l}
\label{transfer}
\end{equation}
where $\psi_0(l)$ denotes the ground-state wave function solving the
Schr\"odinger equation
\begin{equation}
-\frac{1}{2\Sigma}\psi_0(l)''+ U(l)\psi_0(l)=E_0\psi_0(l)
\label{Shrod1}
\end{equation}
with boundary conditions $\psi(0)=\psi(\infty)=0$. We now focus on the
complete wetting limit $\delta\mu\to 0$. Macroscopically far from the apex
the interface unbinds from the wall. Close to the apex however the interface
remains bound due to the pinning exponential term in (\ref{transfer}). 
As $\alpha\to 0$, three different critical behaviors are found:
(I) A MF regime for $p<2$, characterized by Gaussian fluctuations with
$l_A(\alpha)\gg \xi_A(\alpha)$; (II) a marginal case for $p=2$; 
and (III) a fluctuation dominated regime for $p>2$ with universal 
critical behavior and large scale fluctuations $l_A(\alpha)\sim \xi_A(\alpha)
\sim \alpha^{-1}$. The explicit expressions for the large distance/scaling
behavior of $P_A^U(l;\alpha)$, determining the critical singularities, are 
given by:
\begin{equation}
P_A^U(l;\alpha) \sim 
\begin{cases}
l^{\frac{p}{2}}
\exp{(-2\Sigma\alpha l+\frac{4\sqrt{2\Sigma B}}{2-p}l^{1-p/2})} & p<2\\ 
\\
l^{1+\sqrt{1+8\Sigma B}}\exp{(-2\Sigma\alpha l)} & p=2 \\
\\
l^2\exp{(-2\Sigma\alpha l)} & p>2
\end{cases}
\label{MargPDF}
\end{equation}
In the MF regime, a saddle point evaluation reveals that:
\begin{equation}
l_A(\alpha)\sim \left(\frac{2B}{\Sigma \alpha^2}\right)^{1/p};{\indent}
\label{lA}
\end{equation}
and $\nu_A=1/p+1/2$. The critical exponents are continuous at $p=2$.
These results completely classify the asymptotic critical behavior for 
complete wetting at a 2D apex in pure systems. At this point we make 
two remarks:

{\bf{A}}: The values of the critical exponents follow
from a simple mean-field/entropic repulsion argument. Ignoring fluctuations
the equilibrium interfacial profile is obtained 
from minimization of the effective Hamiltonian. A first integral 
of the Euler-Lagrange equation determines the mid-point height at bulk 
coexistence according to
\begin {equation}
\frac{\Sigma \alpha^2}{2}= U(l_A)
\label{MF}
\end{equation}
and leads directly to the result (\ref{lA}) valid in the
MF regime ($p<\tau$). For $p>\tau$ interfacial wandering
leads to an entropic repulsion $U_{fl}\sim l^{-\tau}$. Thus we
should expect two regimes with $\beta_A=\max(2/p,\zeta/(1-\zeta))$
in agreement with the explicit calculation for $\zeta=1/2$. For later
purposes observe that the MF equation (\ref{MF}) is also appropriate for 
higher dimensional apexes.

{\bf{B}}: The PDF's and associated critical exponents are identical to those
occurring at a certain class of 2D critical wetting transition. 
At a critical wetting transition the 
mean height of the interface $l_{\pi}$, roughness $\xi_{\perp}$ and 
parallel correlation length $\xi_{\parallel}$ for a planar substrate
diverge as the temperature (say) is increased towards a wetting 
temperature $T_w$ ($\mu=\mu_{sat}$). This is equivalent 
to the contact angle $\theta$ of a sessile drop vanishing as $T\to T_w^-$. 
The standard interfacial model for this is
\begin{equation}
\beta H_{\pi}[l] = \int dx \left\{ \frac{\Sigma}{2}\left(\frac {dl}{dx}
\right)^2 +V(l)\right\}
\label{effham}
\end{equation}
where $V(l)$ denotes an appropriate binding potential. The associated PDF
$P_{\pi}^V(l;\theta)$ can be calculated using standard methods which map
the problem onto one dimensional quantum mechanics \cite{10}. 
In particular consider 2D critical wetting transitions occurring for 
the class of potentials (\ref{vcov})
with $\chi=\min (p,2)$. For such potentials calculations show that 
$A \propto \theta$.
A straightforward calculation of the PDF's $P_{\pi}^V(l;\theta)$ 
for $p<2$, $p=2$ and $p>2$ yields results that, in the critical
limit (small $\theta$) are \emph{identical} to (\ref{MargPDF}) provided we set
$\theta=\alpha$ \cite{note}. Apex complete wetting precisely mimics the 
properties of a critical wetting transition. 
From the covariance of the PDF's it follows 
that the mean-interfacial heights satisfy
\begin{equation}
l_A(\alpha)=l_{\pi}^V(\alpha)
\label{heightslaw}
\end{equation}
where the RHS is understood to represent the mean height at a planar critical 
wetting transition with the covariant potential (\ref{vcov}).  
It is notable that apex covariance is obeyed at MF level and beyond, and
therefore not necessarily related to fluctuation induced effects such as
hyperscaling. In particular the relation (\ref{heightslaw}) follows directly 
from comparing the solution of the MF equation (\ref{MF}) with the position of
the minimum of the critical wetting potential (\ref{vcov}) (with $\chi=p$).
The covariance relations for apex 
complete wetting are similar, but not identical, to those which exists for 
2D wedge filling transitions. For both pure and impure systems 2D filling 
transitions mimic the properties 
of planar critical wetting transitions with short-ranged forces in contrast 
to the present apex problem where the equivalent critical wetting transition 
has long(er)-ranged forces. Wedge covariance and filling are closely related to
the Indekeu-Robledo conjecture for the line tension \cite{12}, and the
unzipping transition for stranded polymer chains  
\cite{13}. It is likely that similar connections may also apply for 
the apex geometry.

Having discussed the 2D apex in detail it is straightforward to generalise the
results to 3D systems based on the interfacial model
\begin{equation}
\beta H_A[l] = \int d{\mathbf{x}} 
\left\{ \frac{\Sigma}{2} \left(\nabla l\right)^2
+U(l+\alpha\vert x\vert) \right\}
\label{effapex3D}
\end{equation}
where for purposes of generality we have written
${\mathbf{x}}=(x,{\mathbf{x_\parallel}})$ where ${\mathbf{x_\parallel}}$ 
denotes the $d-2$ dimensional vector along the apex axis. In 3D 
${\mathbf{x}}=(x,y)$ but it is instructive to also consider the
generalized apex for $2\le d<3$ since this gives clear indication that the
covariance relations extend to higher dimensions. First
rewrite the Hamiltonian in terms of the relative height $\tilde l$. 
The critical behavior follows from elementary RG considerations. Under
rescaling ${\mathbf{x}}\to{\mathbf{x}}'=
{\mathbf{x}}/b$, $l\to l'=l b^{-\zeta}$ the  
the renormalised tilt angle and Hamaker constant are 
$\alpha'=\alpha b^{1-\zeta}$ and $B'=B b^{2-\zeta p-2\zeta}$ respectively. 
Thus $\alpha$ is always a relevant scaling field whilst the intermolecular 
forces are only relevant for $p<\tau$. The criticality 
falls into two scaling regimes consistent with the explicit 2D 
results:

({\bf{I}}) A MF regime for $p<\tau$ with $\beta_A=2/p$, $\nu_y=2/p+1$ and 
$\nu_A=\zeta \nu_y$ for which (\ref{MF}) is valid.

({\bf{II}}) A fluctuation-regime for $p>\tau$ describing the
universality class of systems with short-ranged forces with
$\beta_A=\zeta/(1-\zeta)$, $\nu_y=1/(1-\zeta)$ and $\nu_A=\zeta
\nu_y$.

Note that for fixed $p$ the upper critical dimension $d_A=3-4/(p+2)$ and 
is unchanged from the planar complete wetting result $d_{co}$. Remarks 
{\bf A} and {\bf B} made earlier about the 2D results also apply in higher 
dimensions, where the planar covariant effective Hamiltonian is:
\begin{equation}
\beta H_{\pi}[l] = \int d{\mathbf{x}} 
\left\{ \frac{\Sigma}{2} \left(\nabla l\right)^2
+V(l) \right\}
\label{effham3D}
\end{equation}

For the physically relevant case $d=3$, MF theory is valid
for all long-ranged intermolecular forces (finite $p$). Thus for non-retarded
van der Waals forces we predict $l_A(\alpha)\sim \sqrt{2B/\Sigma
\alpha^2}$ and observe that this is identical (covariant) with the growth 
of the interfacial thickness at a critical wetting transition with binding 
potential $V=-A/l+B/l^2$. Similar remarks apply for the interfacial 
roughnesses at the respective transitions ($\xi_A(\alpha)\sim \xi_\perp \sim
\sqrt{-\ln \alpha}$). Our final task is to address the 
issue of covariance for the marginal case of 3D systems with short-ranged 
forces. For this we use the interfacial model (\ref{effapex3D}) with 
binding potential $U=B\exp(-\kappa l)$ and $\kappa$ the inverse bulk 
correlation length. At MF level we find $\kappa l_A(\alpha)\sim 
-2 \ln\alpha$ whilst solution of the Ornstein-Zernike equation for the 
height-height correlation function along the apex tip yields 
$\xi_y\sim\alpha^{-1}$ \cite{note2}. Thus the MF exponents for short-range 
forces are $\beta_A=0(\ln)$, $\nu_y=1$ and are consistent with the 
$\zeta\to 0$ limit of the short-ranged exponents detailed in ({\bf{II}}). 
Beyond MF we anticipate that $\nu_y$ is 
unchanged but that the logarithmic divergence of $l_A$ is altered. 
To study this we employ the same approximate linear functional RG 
approach used for the 3D planar wetting transition but with an 
appropriately modified matching condition for the new geometry \cite{14}. 
Under the rescaling of $l$ and ${\mathbf{x}}$ described earlier 
(with $\zeta=0$) the binding potential maps as $U(l)\to R^{(b)}[U]$ 
which effectively coarse-grains the potential over the  
interfacial roughness. Since the angle and correlation length rescale
as $\alpha'=b\alpha$ and $\xi'_y=\xi_y/b$ 
one can curtail the renormalization at $b\sim\alpha^{-1}$ at which scale 
both the renormalized angle and transverse correlation length are of order 
unity and fluctuation effects are negligible. Matching with MF theory implies
$\Sigma/2=R^{(1/\alpha)}[U(l_A)]$ and a simple calculation yields
\begin{equation}
\kappa l_A(\alpha)\sim -(2+\omega)\ln\alpha
\label{RGSR}
\end{equation}
where $\omega=\kappa^2/(4\pi\Sigma)$ is the usual wetting parameter
and we have assumed $\omega<2$ as pertinent to the bulk Ising universality
class. This critical behavior is again consistent with
covariance as can be seen by comparison with the 3D short-ranged critical 
wetting transition described by the model (\ref{effham3D}) with potential
\begin{equation}
V(l)=-A e^{-\frac{\kappa l}{2}} + B e^{-\kappa l}{\indent};l>0
\label{VSR}
\end{equation}
where, guided by our earlier findings, we have included an attractive term
which is twice the range of the direct repulsion. This is the same binding
potential appearing in the standard theory of short-ranged critical wetting 
except for a trivial factor of two in the definition of the inverse bulk 
correlation length. Accordingly, rescaling $\kappa$ and $\omega$ 
in the known RG results for critical wetting  
\cite{14}
\begin{equation}
\kappa l_{\pi}^V(\theta)\sim -(2+\omega)\ln\theta
\label{RGSRV}
\end{equation}
where the (rescaled) wetting parameter $\omega<2$. Note also that the 
divergence of $\xi_y\sim \alpha^{-1}$ is similar to the behaviour of the
critical wetting transverse correlation length \emph{written in terms
of the contact angle} $\xi_\parallel(\theta)\sim \theta^{-1}$. Interestingly 
one still finds the critical behavior (\ref{RGSR}) if one improves the 
apex calculation to account for a position-dependent stiffness coefficient 
\cite{15}. That is, even if planar short-ranged wetting transitions are
driven first-order by a stiffness instability mechanism, the apex 
still mimics the properties of critical wetting.

We finish with comments relevant to experimental studies on periodic systems.
Close to bulk coexistence the interfacial height
above a single, infinite apex shows scaling behavior $l_A \sim 
\alpha^{-\beta_A} W_{(1)}(\delta \mu \alpha^{-\Delta_A})$ with gap exponent
$\Delta_A=2+\beta_A$. On a periodic array (with wavelength $L$) finite size
effects modifies this to $l_A \sim \alpha^{-\beta_A} 
W_{(2)}(\delta \mu \alpha^{-\Delta_A},L/\xi_\parallel^{co})$, where 
$\xi_\parallel^{co}\sim \delta \mu^{-\nu_\parallel^{co}}$ is the transverse
correlation length for planar complete wetting. Thus the adsorption 
above the apex tip only behaves like a single apex for sufficiently
large $L\gg \xi_\parallel^{co}$. This is equivalent to the requirement 
that the vertical distance between the apex tip and the wedge trough,
$\alpha L /2$, is much larger that the local height of the interface above
the wedge bottom $l_w \sim \Sigma \alpha^2 / 2 \beta \delta \mu 
(\rho_l - \rho_g)$.

In this paper we have shown that complete wetting at an apex mimics precisely
planar critical wetting. Taken together with similar covariance relations 
for wedge filling, there is clear evidence of a fundamental 
connection between contact and geometric angles. Further work is 
required to understand such covariances at a deeper level.

J.M.R.-E. and M.J.G. acknowledge financial support from Secretar\'{\i}a de 
Estado de Educaci\'on y Universidades (Spain), co-financed by the 
European Social Fund, and EPSRC (UK), respectively.


\begin{thebibliography}{199}
\bibitem{1} S. Dietrich, in {\it New approaches to Old and New Problems
in Liquid-State Theory}, edited by C. Caccamo \emph{et al}, 
NATO-ASI, Ser. B (Kluwer, Dordrecht, 1998), p 197.
\bibitem{2} H. Gau \emph{et al}, Science {\bf
283}, 46 (1999).
\bibitem{3} C. Rasc\'on and A. O. Parry, Nature {\bf 407}, 986 
(2000). 
\bibitem{4} S. Gheorghiu and P. Pfeifer, Phys. Rev. Lett. {\bf 85}, 
3894 (2000).
\bibitem{5} L. Bruschi, A. Carlin and G. Mistura, J. Chem. Phys. {\bf 115},
6200 (2000); Phys. Rev. Lett. {\bf 89}, 166101 (2002).
\bibitem{6} For a general review of wetting, see for example S. Dietrich,
in {\it Phase Transitions and Critical Phenomena}, vol 12, edited by C. Domb
and J. L. Lebowitz (Academic Press, New York, 1988), p 1. 
\bibitem{7} For recent experimental reviews, see 
D. Bonn and D. Ross, Rep. Prog. Phys. {\bf 64}, 1085 (2001), and 
B. M. Law, Prog. Surf. Sci. {\bf 66}, 159 (2001). 
\bibitem{8} M. E. Fisher, J. Chem. Soc. Faraday Trans. 2 {\bf 82}, 1589 
(1986).
\bibitem{9} P. Huber \emph{et al} Phys. Rev. Lett. {\bf 89}, 035502 (2002).
\bibitem{10} R. Lipowsky, Phys. Rev. B {\bf 32}, 1731 
(1985) 
\bibitem{11} 
A. O. Parry, M. J. Greenall and A. J. Wood, J. Phys.: Condens. Matter {\bf 
14}, 1169 (2002).
\bibitem{note} The exact covariant potential $V=-\alpha \psi'_0/\psi_0 + U$
reduces to the Eq. (\ref{vcov}) for large $l$. 
\bibitem{12} J. O. Indekeu and A. Robledo, Phys. Rev. E {\bf 47}, 4607 
(1993).
\bibitem{13} D. K. Lubensky and D. R. Nelson, Phys. Rev. Lett. {\bf 85},
1572 (2000).
\bibitem{note2} Further details will be published elsewhere.
\bibitem{14} D. S. Fisher and D. A. Huse, Phys. Rev. B {\bf 32}, 247
(1985). 
\bibitem{15} M. E. Fisher and A. J. Jin, Phys. Rev. Lett. {\bf 69}, 792 
(1992).
\end{thebibliography}
\end{document}